\documentstyle[sprocl]{article}

\input{psfig}

\def\Journal#1#2#3#4{{#1} {\bf #2}, #3 (#4)}

\def\NPB{{\em Nucl. Phys.} B}
\def\PLB{{\em Phys. Lett.}  B}
\def\PRL{\em Phys. Rev. Lett.}

\def\ZPC{{\em Z. Phys.} C}

\def\be{\begin{equation}}
\def\ee{\end{equation}}
\def\bea{\begin{eqnarray}}
\def\eea{\end{eqnarray}}
\def\etal{et al.}
\begin{document}

\title{PATHOLOGICAL SCIENCE}

\author{SHELDON STONE}

\address{Physics Department, Syracuse University, Syracuse\\
 NY 13244-1130, USA\\E-mail: stone@phy.syr.edu}


\maketitle\abstracts{I discuss examples of what Dr. Irving Langmuir, a Nobel
prize winner in Chemistry, called ``the science of things that aren't so."  
Some of his examples are reviewed and others from High Energy
Physics are added. It is hoped that discussing these incidents will help us
develop an understanding of some potential pitfalls.}

\section{Introduction}
Often, much more often than we would like, experimental results are reported
that have impressive ``statistical significance," but are subsequently proven to
be wrong. These results have labeled by Irving Langmuir as the {\it ``Science
of things that aren't so."} Langmuir described some of these incidents in
a 1953 talk that was transcribed and may still be available.\cite{Langmuir}
Here I will repeat some of Langmuir's examples, and show some other examples
from High Energy Physics.

The examples shown here are not cases of fraud; the proponents believed in the
work they presented. But they were wrong.

\section{Davis-Barnes Effect}

\begin{figure}[htb]
%\rule{5cm}{0.2mm}\hfill\rule{5cm}{0.2mm}
%\vskip 2.5cm
%\rule{5cm}{0.2mm}\hfill\rule{5cm}{0.2mm}
\centerline{\psfig{figure=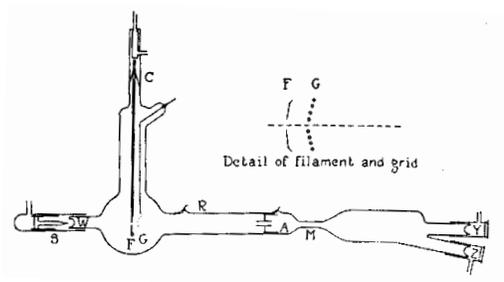,height=1.5in,angle=-1.5}}
\caption{Diagram of first tube. S, radioactive source; W, thin glass window;
F, filament; G, grid; R, lead to silvered surface; A, second anode;
M, magnetic field; C, copper seals; Y and Z, zinc sulfide screens. 
\label{fig:bergen_davis1}}
\end{figure}

Circa 1930 Professors B. Davis and A. Barnes of Columbia University did an
experiment where they produced $\alpha$ particles from the decay of Polonium
and $e^-$ from a filament in an apparatus sketched in Fig.~\ref{fig:bergen_davis1}. The
electrons are accelerated by a varying potential. At 590 V they
move with the
same velocity as the $\alpha$'s. Then they may combine with the $\alpha$'s to
form a bound $\alpha-e^-$ ``atomic state." They then continue down the tube and
are counted visually by making scintillations in the screens at Y or Z that
viewed with a microscope.

Without a magnetic field all the $\alpha$'s reach the screen at Y. With
a magnetic field and no accelerating voltage for the electrons they all reach the screen at
Z. However, if the electrons bind with the doubly positive charged $\alpha$'s
they would deflect half as much and not reach Z. 

\begin{figure}[htb]
\centerline{\psfig{figure=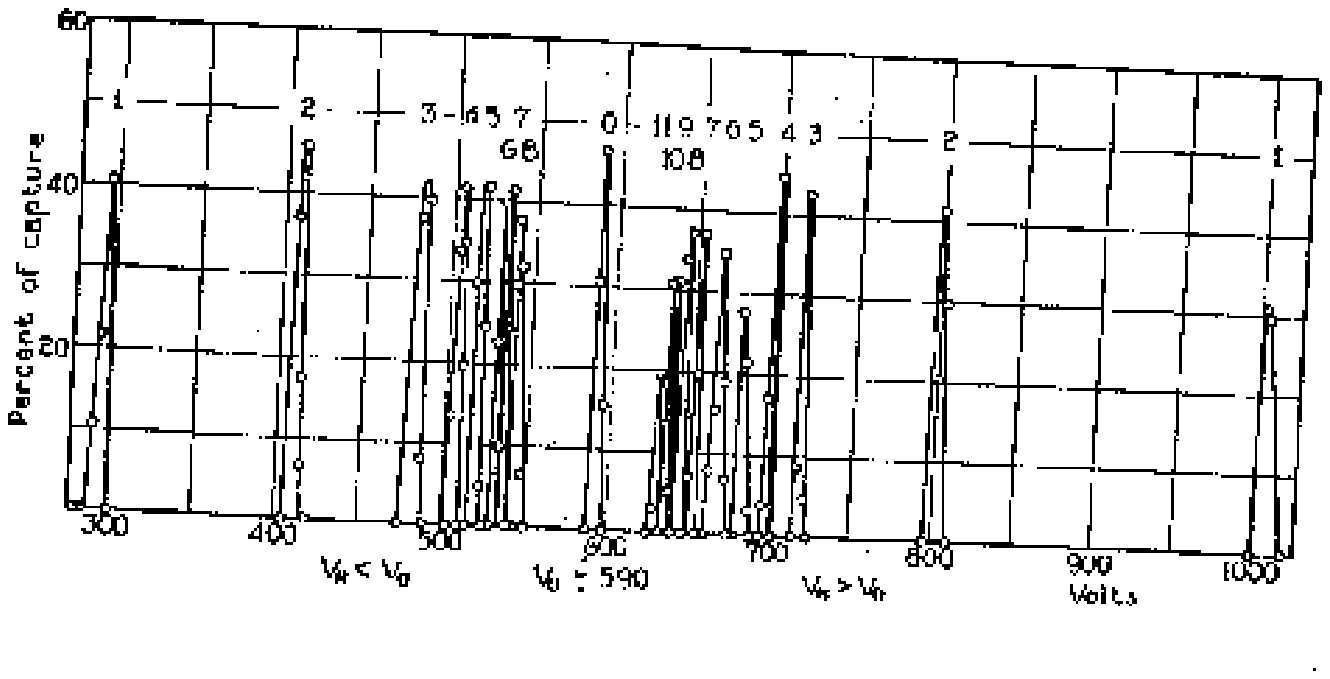,height=2.0in,angle=3}}
\vspace{-7mm}
\caption{Electron capture as a function of accelerating voltage. 
\label{fig:bergen_davis2}}
\end{figure}

What they found was very extraordinary. Not only did the electrons combine with
the $\alpha$'s at 590 V, but also at other energies ``that were exactly the
velocities that you calculate from Bohr theory." Furthermore all the capture
probabilities were about 80\%. Their data are shown in
Fig.~\ref{fig:bergen_davis2}.

Now there is a problem here because in Bohr theory when an electron comes in
from infinity it has to radiate half its energy to enter into orbit. There was
no evidence for any such radiation and the electron would have needed to have
twice the energy to start with. However, there were some theorists including
Sommerfeld who had an explanation that the electron could be captured if it had
a velocity equal to what it was going to have in orbit.

There were other disturbing facts. The peaks were 0.01 V wide; the fields in
the tube were not that accurate. In addition, scanning the entire voltage range
in such small steps would take a long time. Well, Davis and Barnes explained,
they didn't quite do it that way: they found by some preliminary work that they
did check with the Bohr orbit velocities, so they knew where to look. Sometimes
they weren't quite in the right place, so they explored around and found them.
Their precision was so good they were sure they could get a better value for
the Rydberg constant (known then to 1 part in 10$^8$).

Then Langmuir visited Columbia. The way the experiment was done was that an
assistant named Hull sat opposite to Barnes in front of a voltmeter,
that had a scale that went from 1 to a thousand volts and on that scale he was
reading 0.01 V. The room was dark, to see the scintillations, and there was a
light on the voltmeter and on the dial of the clock that Barnes used to time
his measurements. 

Langmuir says it best: ``He said he always counted for two minutes. Actually, I
had a stop watch and I checked him up. They sometimes were as low as one minute
and ten seconds and sometimes one minute and fifty-five seconds, but he counted
them all as two minutes, and yet the results were of high accuracy!

``And then I played a dirty trick. I wrote out on a card of paper ten different
sequences of V and zero. I meant to put on a certain voltage and then take it
off again. Later I realized that that wasn't quite right because when Hull took
off the voltage, he sat back in his chair---there was nothing to regulate at
zero, so he didn't. Well, of course, Barnes saw him whenever he sat back in
his chair. Although, the light wasn't very bright, he could see whether he was
sitting back in his chair or not so he knew the voltage wasn't on and the
result was that he got a corresponding result. So later I whispered, `Don't let
him know that you're not reading,' and I asked him to change the voltage from
325 down to 320 V so he'd have something to regulate and I said, `regulate it
just as carefully as if you were sitting on a peak.' So he played the part from
that time on, and from that time on Barnes' readings had nothing whatever to do
with the voltages that were applied. Whether the voltage was at one value or
another didn't make the slightest difference.  I said `you're through. You're
not measuring anything at all. You never {\it have} measured anything at all.'
`Well,' he said, `the tube was gassy. The temperature has changed and
therefore the nickel plates must have deformed themselves so that the
electrodes are no longer lined up properly.'

``He immediately---without giving any thought to it---he immediately had an
excuse. He had a reason for not paying any attention to any wrong results. It
just was built into him. He just had worked that way all along and always
would. There is no question but what he is honest; he {\it believed} these
things, absolutely."

In fact they did publish their results even after being confronted by
Langmuir.\cite{bergen_davis} Later after no one else had been able to reproduce their
results they published a retraction,\cite{bdr} that said in part: ``These
results reported depended on observations made by counting scintillations
visually. The scintillations produced by $\alpha$ particles on a zinc sulfide
screen are a threshold phenomenon. It is possible that the number of counts may
be influenced by external suggestion or autosuggestion to the observer. The
possibility that the number of counts might be greatly influenced by suggestion
had been realized, and a test of their reliability had been made by two
methods: (a) The voltage applied to the electrons was altered without the
knowledge of the observer (Barnes); (b) the direction of the electron stream
with respect to the $\alpha$-particle path was altered by a small
electro-magnet. Such changes in voltage and direction of electron stream were
noted at once by the observer. These checks were thought at the time to be
entirely adequate. In examining the data of observation made in our laboratory
Dr. Irving Langmuir concluded that the checks applied had not been sufficient,
and convinced us that the experiments should be repeated by wholly objective
methods. Accordingly we have investigated the matter by means of the Geiger
counter. Four additional experimental electron $\alpha$-ray tubes have been
constructed for this purpose.

``Capture of the kind reported was often observed over a considerable period of
time, but following prolonged observation the effect seemed to disappear. The
results deduced from visual observations have not been confirmed. If such
capture of electrons does take place, it must depend on unknown critical
conditions which we were not able to reproduce at will in the new experimental
tubes."

It is interesting to note that
they still seemed to be holding out the possibility that somehow the earlier
results were correct.

\section{N-rays}

In 1903 there was a lot of experimentation with x-rays. Blondlot, a respected
member of the French Academy of Sciences, found that if you have a hot wire heated
inside an iron tube with a window cut out of it, rays would emerge that would
get through aluminum. He call these N-rays.\cite{Blondlot} They had specific
properties. For example, they could get
through 2" or 3" of aluminum but not through iron. The way he detected these
rays was by observing an object illuminated with a faint light. When the N-rays
were present you could ``see the object {\it much} better."

N-rays could be stored. Brick wrapped in black paper put in sunlight would
store and reemit them, but the effect was independent of the number of bricks.
Many other things would give off N-rays, including people. They even split when
traversing an aluminum prism. Blondlot measured the index of refraction of the
different components.

The American physicist R. W. Wood visited Blondlot's laboratory and was shown
the experiments. While Blondlot demonstrated his measurement of the refractive
indicies, Wood palmed the prism. It did not affect the measurements. Wood
cruelly published that,\cite{Wood} and that was the end of Blondlot.

Now the question is how do we explain Blondlot's findings. Pringsheim tried to
repeat Blondlot's experiments and focused on the detection method. He found
that if you have a very faint source of light on a screen of paper and to make
sure that you are seeing the screen of paper you hold your hand up and move it
back and forth. And if you can see your hand move then you know it is
illuminated. One of Blondlot's observations  was that you can see much
better if you had some N-rays falling on the piece of paper. Pringsheim
repeated these and found that if you didn't know where the paper was, whether
it was in front or behind your hand, it worked just as well. That is you could
see your hand just as well if you held it back of the paper as if you held it
in front. Which is the natural thing, because this is a threshold phenomenon,
and a threshold phenomenon means that you don't know, you really don't know,
whether you are seeing it or not. But if you have your hand there, well of
course, you see your hand because you {\it know} your hand's there, and that's 
just enough to win you over to where you know that you see it. But you know it
just as well if the paper happens to be in front of your hand instead of in
back of your hand, because you don't know where the paper is but you {\it do}
know where your hand is.

\section{Symptoms of Pathological Science}
Langmuir lists six characteristics of pathological science:
\begin{enumerate}
\item The maximum effect that is observed is produced by a causative agent of
barely detectable intensity, and the magnitude of the effect is substantially
independent of the cause.
\item The effect is of a magnitude that remains close to the limit of
detectability, or many measurements are necessary because of the low
statistical significance of the results.
\item Claims of great accuracy.
\item Fantastic theories contrary to experience.
\item Criticisms are met by ad hoc excuses thought up on the spur of the
moment.
\item Ratio of supporters to critics rises up to somewhere near 50\% and then
falls gradually to oblivion. 
\end{enumerate}

Let us go on to examples from high energy physics and see how these criteria
apply. Unfortunately, there are many examples. I have just chosen a few.

\section{The Split $A_2$ Meson Resonance}
\begin{figure}[htb]
\centerline{\psfig{figure=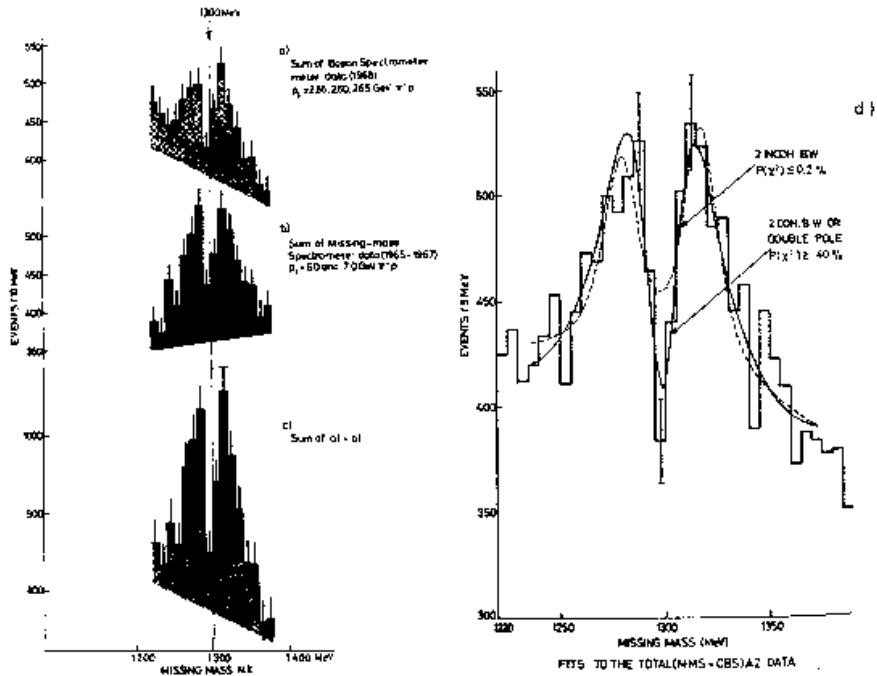,height=3.7in,angle=0}}
\vspace{-2mm}
\caption{(a-c) Evidence for $A_2$ splitting in $\pi^- p\to p X^-$ collisions
in the two CERN experiments, (d) same as (c) in 5 MeV bins fit to two
hypotheses. 
\label{fig:a2_split}}
\end{figure}

Two CERN experiments, the Missing Mass Spectrometer (MMS) experiment,\cite{mms}
and the CERN Boson Spectrometer (CBS) experiment,\cite{cbs} claimed that the
structure around 1300 MeV, believed to be the $2^+$ $A_2$ meson produced in pion proton collisions did not have a
simple Breit-Wigner form, as expected for a short lived resonance, but was in fact divided (or split) into two peaks.
Their data are shown in Fig.~\ref{fig:a2_split}. 

\begin{figure}[htb]
\vspace{-4mm}
\centerline{\psfig{figure=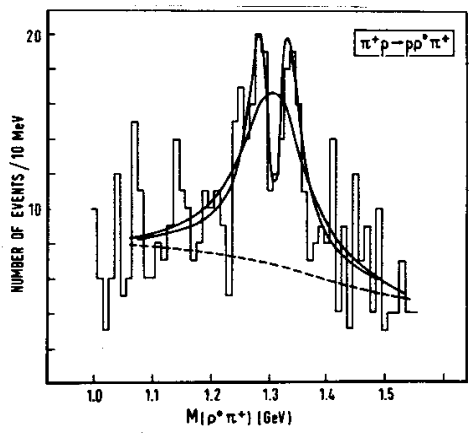,height=2.5in,angle=0}}
\vspace{-4mm}
\caption{Data from B\"ockman \etal~ The simple Breit-Wigner has a 20\%
probability while the double-pole fit gives 63\%. 
\label{fig:Bockman}}
\end{figure}

\begin{figure}[hbt]
\vspace{-2mm}
\centerline{\psfig{figure=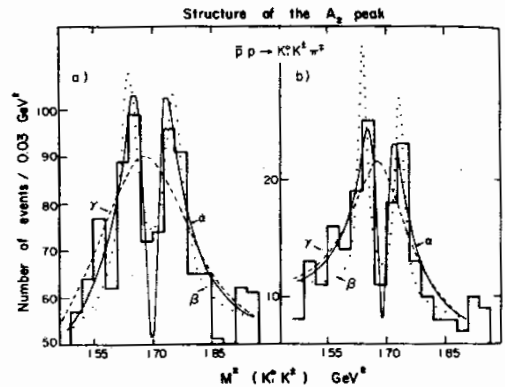,height=2.2in,angle=1}}
\vspace{-6mm}
\caption{Data of Aguilar-Benitez \etal~(a) All $\overline{p}$ momenta
included (note the suppressed zero). (b) Only the 0.7 GeV/c data. Using (a)
only, the fit for double-pole ($\alpha$) gives 65\% likelihood. 
\label{fig:AB}}
\end{figure}

This indeed was a startling result with no obvious explanation. Such a
resonance shape could mean new physics. The MMS experiment just observed the
outgoing proton and thus computed the missing mass from proton and the
knowledge of the incident $\pi^-$ beam. The CBS experiment could also observe
the decay products of the $A_2$. Bubble chamber experiments
also saw evidence for the splitting. B\"ockman \etal~had 5 GeV/c $\pi^+ p$
data.\cite{Bock} They looked at the $\rho^{\circ}\pi^+$ final state and showed their data for a
specific cut on the momentum transfer between the initial and final state proton
($t-t_{min}<0.1$). Their data shows a split (see Fig.~\ref{fig:Bockman}). 
Anguilar-Benitez \etal~\cite{AB} showed their $K^{\circ}K^{\pm}$ data which
they claim fits best to a double pole (see Fig.~\ref{fig:AB}). 
There were also inconclusive but split-suggestive data
from Crennell \etal\cite{Crennell}

\begin{figure}[htb]
\centerline{\psfig{figure=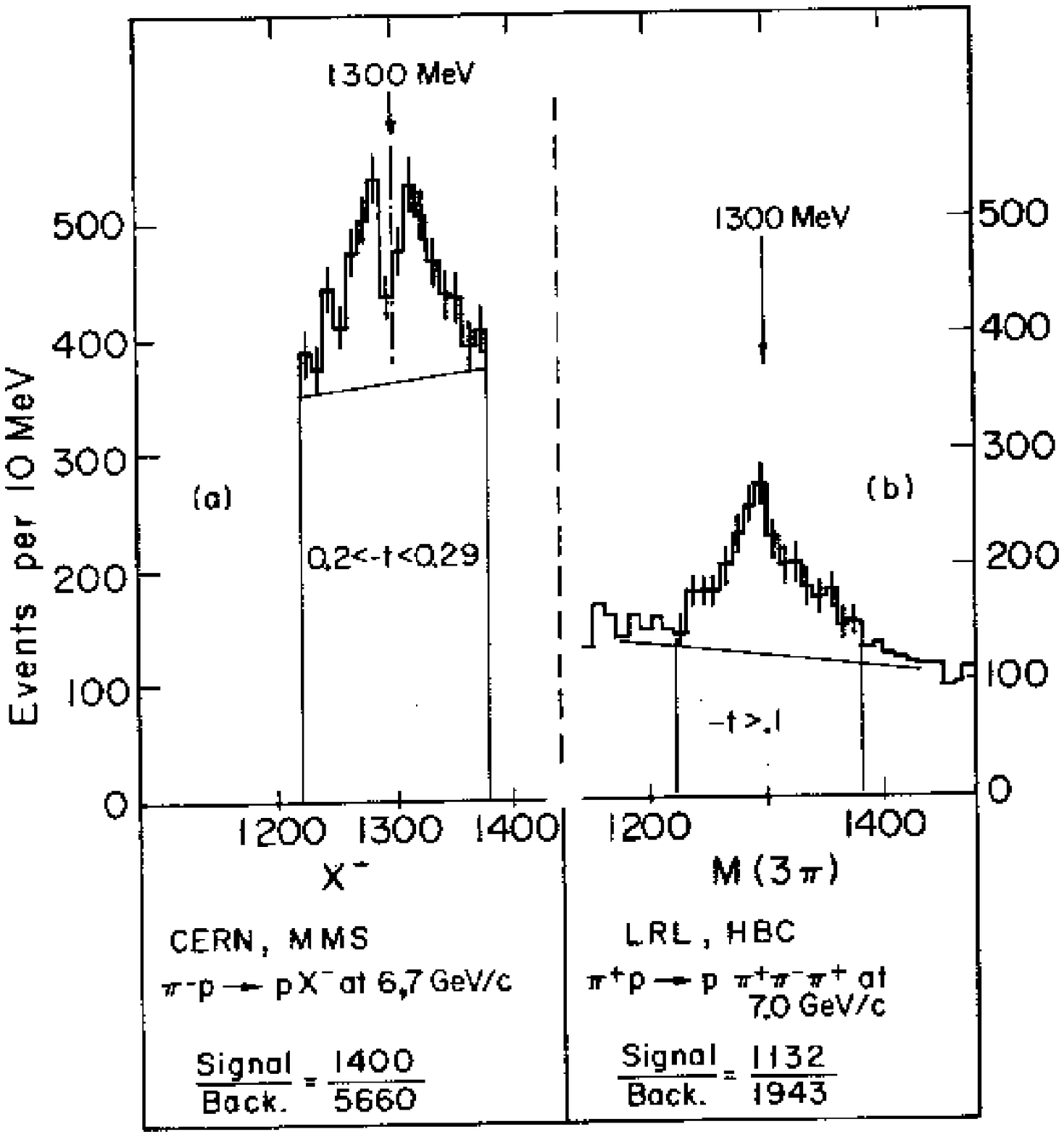,height=3.5in,angle=-1}}
\vspace{-3mm}
\caption{Comparison of MMS and LRL data at similar incident pion momenta.
(a) MMS, resolution $\Gamma$/2=8.0 MeV, (b) LRL, resolution $\Gamma$/2=6.4 MeV. 
\label{fig:lbl}}
\end{figure}

A. Barbaro-Galtieri reviewed the situation at the 1970 Meson Spectroscopy
conference.\cite{Lina} At the time there was only one {\it publicly
available} result that directly contradicted the split. These data, from a
$\pi^+p$ bubble chamber experiment done by the Berkeley group (LRL), are
compared with the MMS data in Fig.~\ref{fig:lbl}.\cite{LRL}

At this conference doubt was raised about the validity of the split. Others
then came forward.\cite{nosplit} There were new experiments.\cite{nonosplit}
By the 1972 Meson Spectroscopy
conference, there was no mention of the split. It had vanished into
oblivion.\cite{meson72}

How did this happen? I have heard several possible explanations. In the MMS
experiment, I was told that they adjusted the beam energy so the dip always
lined up! Another possibility was revealed in a conversation I had with
Sch\"ubelin, one of the CBS physicists. He said: ``The dip was a clear feature.
Whenever we didn't see the dip during a run we checked the apparatus and always
found something wrong." I then asked him if they checked the apparatus when
they did see the dip, and he didn't answer.

What about the other experiments that did see the dip?  Well there were several
experiments that didn't see it. Most people who didn't see it had less
statistics or poorer resolution than the CERN experiments, so they just kept
quiet. Those that had a small fluctuation toward a dip worked on it until it
was publishable; they looked at different decay modes or $t$ intervals, etc.
(This is my guess.) 

\section{The R, S, T and U Bosons}
\begin{figure}[htbp]
\vspace{-5mm}
\centerline{\psfig{figure=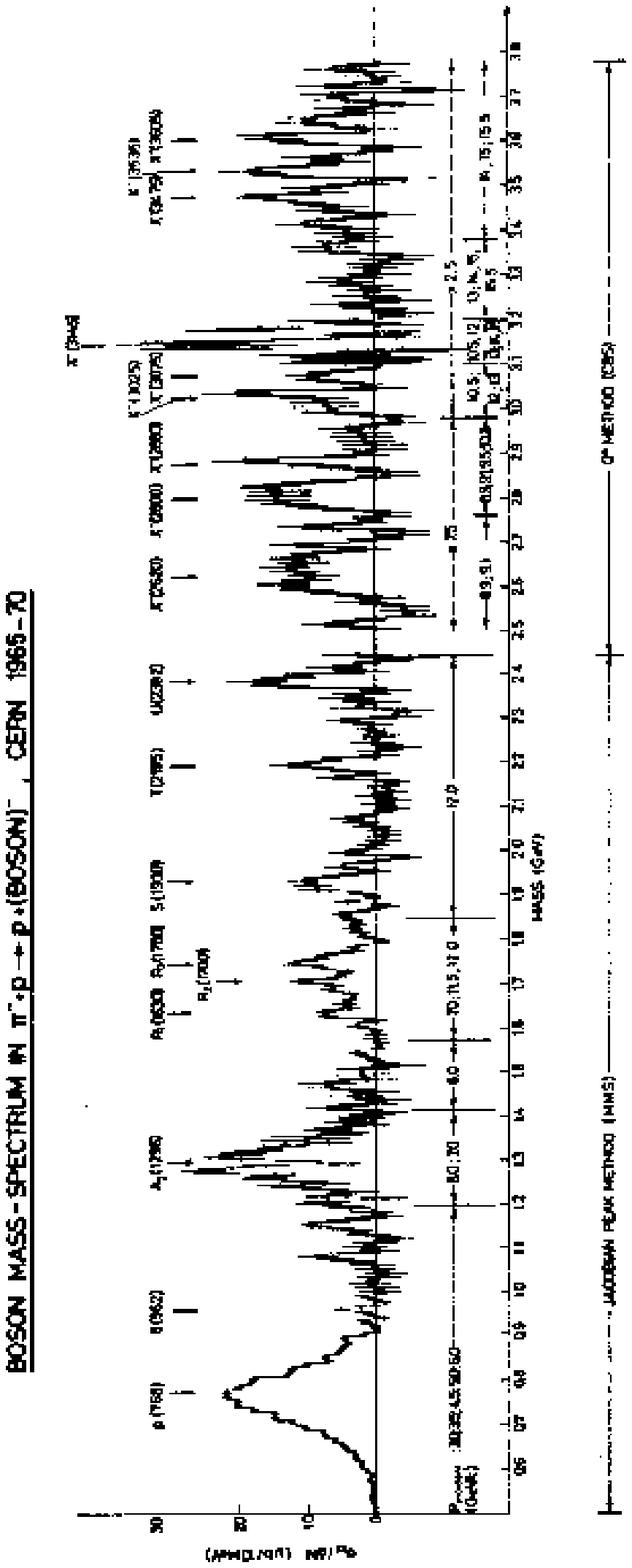,height=6.5in,angle=0}}
\vspace{-5mm}
\caption{Compiled spectrum obtained by adding the spectra in different mass
regions with the hand-drawn background subtracted. The arrows indicate the
position of known or suspected particles.
\label{fig:cmm}}
\end{figure}

The 1970 Meson Spectroscopy conference contained many new results
involving claims for states of relatively narrow width in the 3 GeV mass
region.
The CBS group also presented evidence of significant peaks, more than four
standard deviations for six new resonances above 2.5 GeV/c. Their data are
shown in Fig.~\ref{fig:cmm}.\cite{Damg}

Other groups also saw bumps. At the 1970 conference Miller reported similar structures seen in a 13
GeV/c incident momenta $\pi^+p$ bubble chamber experiment.\cite{Miller}
Kalbfleisch gave a review entitled ``The T Region," in which he stated ``In
this reiview paper I will discuss evidence for mesons in the T region (mass 2.19
GeV). The T$^-$ is {\it well known} from the original missing mass spectrometer work at
CERN."\cite{Kalb}

Subsequently, there were no further confirmations of these signals. In fact,
all of these bumps eventually went away. 

\section{The $F$ or $D_s$ Meson}
The spin-0 meson formed from $c\overline{s}$ constituent quarks was called at first
the $F$, but was renamed by the Particle Data Group as the
$D_s$. The name was not changed to protect the innocent.\cite{dragnet} 
The $F$ decays mostly by having the $c$ quark transform to an $s$ quark and
a virtual $W^+$ boson. In the simplest case, the $W^+$ manifests itself as a
$\pi^+$ and the $s$ quark combines with the original $\overline{s}$ quark to
form a $\phi$ or $\eta$ meson.

In 1977 the DASP group working at the DORIS $e^+e^-$ storage ring at DESY found
a handful of events at a center-of-mass energy of 4.42 GeV that they 
classified as coming from the reaction $e^+e^-\to F^+F^-\gamma$, where
the $\gamma$ and one of the $F$'s formed an $F^*$, the spin-1 state. One $F^{\pm}$ candidate
was required to decay into an $\eta\pi^{\pm}$, while the other $F$ was not
reconstructed.\cite{DASP} The $\eta\to\gamma\gamma$ channel was used.

\begin{figure}[htb]
\vspace{-2mm}
\centerline{\psfig{figure=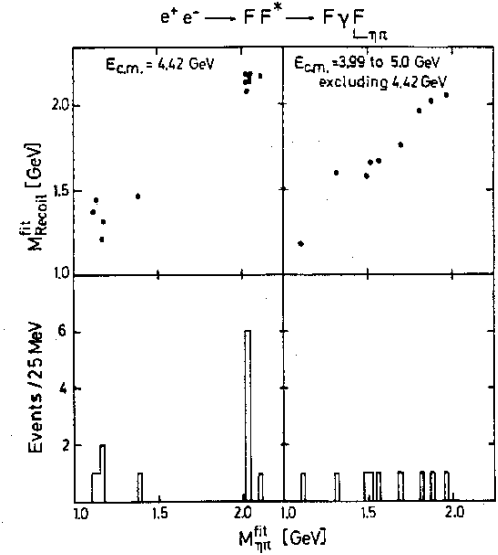,height=3.0in,angle=0}}
\vspace{-2mm}
\caption{(Top) Fitted $\eta\pi^{\pm}$ mass versus fitted recoil mass assuming
$e^+e^-\to FF^*$, where $F^*\to F\gamma$, and $F^{\pm}\eta\pi^{\pm}$ at
(left) $E_{cm}$=4.42 GeV and (right) at all other energies excluding 4.42 GeV.
At the bottom are the fitted $M(\eta\pi)$ projections. 
\label{fig:dasp}}
\end{figure}

The data were fit to the $FF^*$ hypothesis requiring that both the $F^+$ and
the $F^-$ have the same mass.  Their results are shown in Fig.~\ref{fig:dasp}.
At a center of mass energy of 4.42 GeV they observe a cluster of events in
$\eta\pi^{\pm}$ mass above 2 GeV, and no such cluster at other energies. 
They also observed an increase in the production of $\eta$ mesons
in 4.42 GeV region. (More on this later.)
The $F$ mass had an ambiguity because the low momentum photon from the $F^*$
decay could be associated either with the $F$ that decayed into $\eta\pi$ or
with the $F$ that wasn't reconstructed. Thus they found
two possible mass values for the $F$, 2040$\pm$10 MeV or 2000$\pm$40 MeV.
The generally quoted value was 2020-2030 MeV.

\begin{figure}[htb]
\centerline{\psfig{figure=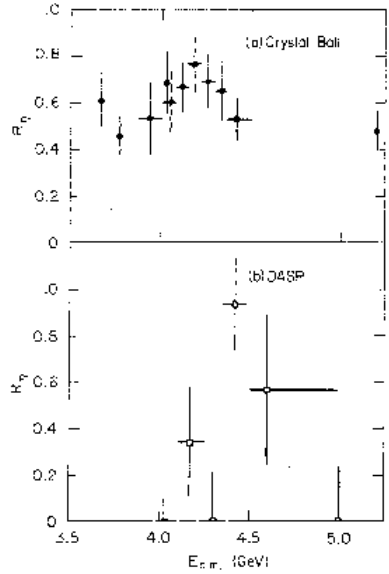,height=3.3in,angle=0}}
\vspace{-6mm}
\caption{$R_{\eta}=\sigma(e^+e^-\to\eta X)/\sigma(e^+e^-\to\mu^+\mu^-)$ for
(a) Crystal Ball and (b) DASP. 
\label{fig:eta_prod}}
\end{figure}

There was however a disturbing aspect of this result. The
Crystal Ball group, operating at SPEAR checked the level of $\eta$ production
in the same center-of-mass energy region.\cite{CB} A comparison of their result
with the DASP result is shown in Fig.~\ref{fig:eta_prod}. 
Crystal Ball, which had a far superior ability to detect photons compared to
DASP, did not find the increase in $\eta$ production at 4.42 GeV that DASP
claimed. Without an increase in $\eta$ production it's hard to see why $FF^*$
would be produced at 4.42 GeV and not at other energies. Unfortunately the Crystal Ball experiment did
not have charged track momentum analysis and could not look for $F$'s.

\begin{figure}[htb]
\centerline{\psfig{figure=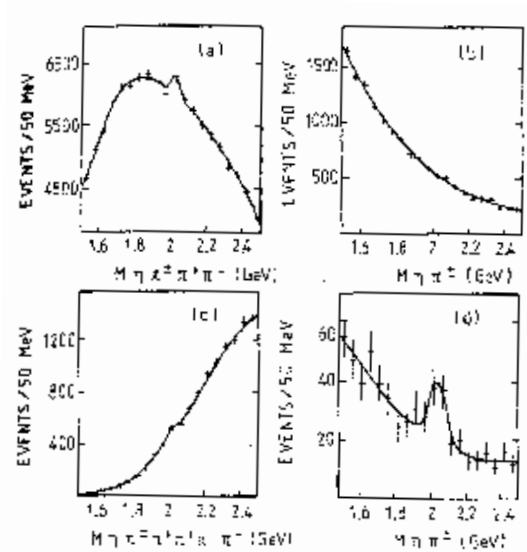,height=3.0in,angle=-1}}
\vspace{-3mm}
\caption{Mass spectra of (a) $\eta\pi^{\pm}\pi^+\pi^-$ from T1, (b) 
$\eta\pi^{\pm}$ from T1, (c) $\eta\pi^{\pm}\pi^+\pi^-\pi^+\pi^-$ from T1, and
(d) $\eta\pi^{\pm}$ from T2. The curves are polynomial plus
Breit-Wigner fits.  
\label{fig:omega1}}
\end{figure}

In 1981 a group using the CERN Omega Spectrometer with a 20-70 GeV photon beam
found evidence for the $F$
meson at 2020$\pm$10 MeV in several different decay modes. They used two sets of
selection criteria to record data, denoted T1 and T2.\cite{omega_1} T1 required a minimum of
four charged particles at a plane 1.5 m downstream of the target center. T2
required a photon with transverse momentum greater than 800 MeV and at least
one charged track leaving the target. They show results for selected decay
modes containing $\eta\to\gamma\gamma$ in
Fig.~\ref{fig:omega1}. 
\begin{figure}[htb]
\vspace{-5mm}
\centerline{\psfig{figure=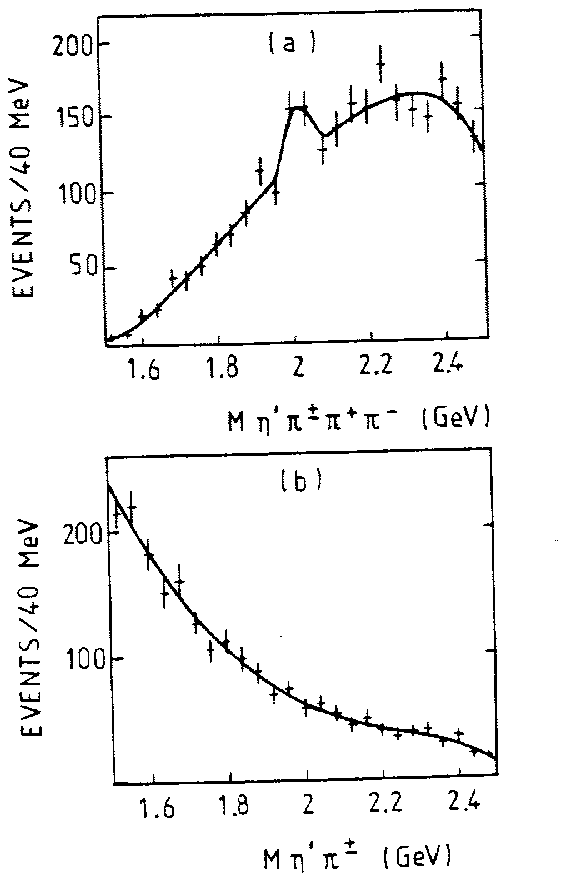,height=3.0in,angle=-1}}
\vspace{-5mm}
\caption{Mass spectra of (a) $\eta'\pi^+\pi^+\pi^-$, a subset of (c) 
in Fig.~\ref{fig:omega1}, (b) 
$\eta\pi^{\pm}$, a subset of (a) in Fig.~\ref{fig:omega2}. The curves are polynomial plus
Breit-Wigner fits.  
\label{fig:omega2}}
\end{figure}

In Fig.~\ref{fig:omega2} they select candidates for $\eta'\to
\pi^+\pi^-\eta$ from the $\eta 5\pi$ and $\eta 3\pi$ samples. They see a signal
in $\eta'3\pi$ and nothing in $\eta'\pi$.

\begin{figure}[htb]
\centerline{\psfig{figure=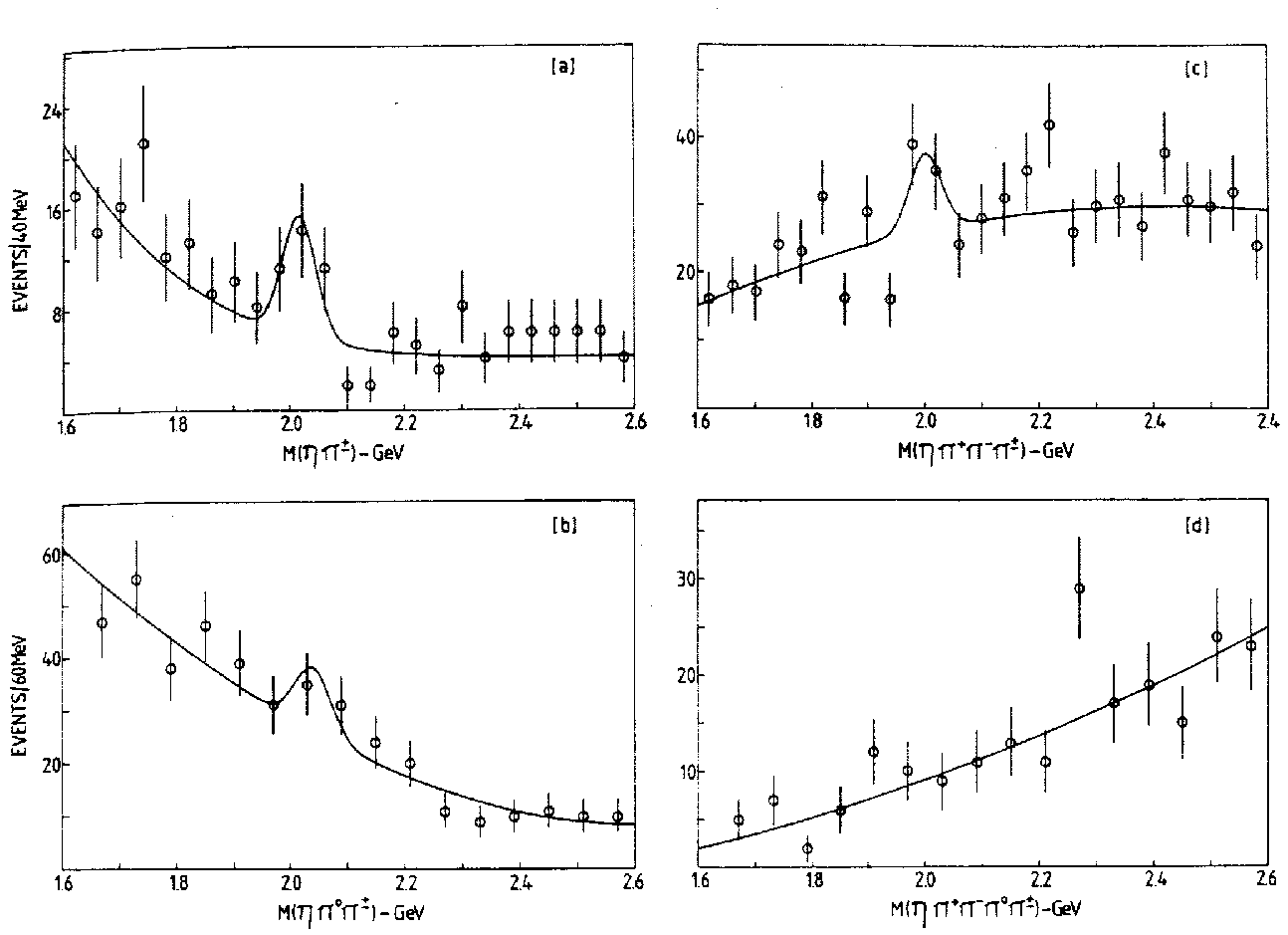,height=3.0in,angle=0}}
%\vspace{-7mm}
\caption{Mass spectra of  $\eta(n\pi^{\pm})$ from the CERN Omega spectrometer.
The curves are polynomial plus Gaussian fits.  
\label{fig:omega3}}
\end{figure}

In a 1983 paper they presented results based on a different trigger where they
required one photon with energy greater than 2 GeV and forward charged
multiplicity between 2 and 5.\cite{omega_2} Their data are shown in Fig.~\ref{fig:omega3}.
The $F$ mass value was virtually unchanged.

These newer signals are quite weak, as was a signal they found in
$\phi\rho^{\pm}$. They did not see a signal in $\phi\pi^{\pm}$.\cite{omega_3}
For the $\phi$ results they used a different trigger that required between
4 and 9 charged tracks including an identified $K^{\pm}$, called T4.
This group also published results on $D$ meson production.\cite{omega_4} The relevant
yields are listed in Table~\ref{tab:omega}.
\begin{table}[htbp]
\caption{CERN Omega Spectrometer Photo-production Yields for $F$ Mesons into
$\eta$, $\eta'$ or $\phi$ and $D$ Modes into Kaons.
\label{tab:omega}}
\vspace{0.2cm}
\begin{center}
\footnotesize
\begin{tabular}{|l|r|r|r|r|}
\hline
\raisebox{0pt}[13pt][7pt]{Mode} &
\raisebox{0pt}[13pt][7pt]{Trigger} &
\raisebox{0pt}[13pt][7pt]{Efficiency} &
\raisebox{0pt}[13pt][7pt]{\# of events or significance}&
\raisebox{0pt}[13pt][7pt]{$\cal{B}\times\sigma$ (nb)}\\
\hline
$\eta\pi^+\pi^-\pi^{\pm}$ & T1 & 10 &4$\sigma$ & 60$\pm$15\\
$\eta'\pi^+\pi^-\pi^{\pm}$ & T1 & 5 &3$\sigma$ &20$\pm$~~8\\
$\eta\pi^{\pm}$ & T2 & 3 & 5$\sigma$ & 27$\pm$~~7\\
$\eta\pi^{\pm}$ & T1 & 12 & &$<$45\\
$\eta'\pi^{\pm}$ & T1 & 6 & &$<$30\\\hline
$\eta\pi^{\pm}$ & T3 & 1.83 & 17$\pm$6& 38$\pm$14\\
$\eta\pi^{\circ}\pi^{\pm}$ & T3 & 0.85 &14$\pm$9& 66$\pm$42\\
$\eta\pi^+\pi^-\pi^{\pm}$ & T3 & 0.87 &20$\pm$11& 93$\pm$52\\\hline
$\phi\pi^{\pm}$& T4 & & & $<$4\\
$\phi\rho^{\pm}$& T4 & &3$\sigma$ & 33$\pm$10\\
$\phi\pi^+\pi^-\pi^{\pm}$& T4 & & & $<$15\\\hline
$K^-\pi^+$ & & 42 & 60$\pm$17 & 13.5$\pm$~4\\
$K^-\pi^+\pi^{\circ}$ & & 5 & 63$\pm$19 &108$\pm$33\\
$K^{\circ}\pi^+\pi^-$ & & 17 & 66$\pm$19&39$\pm$11\\\hline
\multicolumn{5}{l}{Efficiencies include $\eta^{(')}$ decay fractions. Upper
limits are at 3$\sigma$.}

\end{tabular}
\end{center}
\end{table}

Now there are several significant problems here comparing $D^{\circ}$ production with
$F^{\pm}$ production. First of all product of the branching ratio times
production cross section ($\sigma\cdot\cal{B}$) for $F^{\pm}\to \eta\pi^{\pm}$ is more than twice as
large as for $D^{\circ}\to K^-\pi^+$. It is expected that the $F$, being a
charmed-strange meson, would be produced approximately 15\% as often as the
charmed-light quark combination. While the two-body branching ratios cannot be
accurately predicted, the $\eta\pi$ would have to be about 14 times larger in the $F$ than
the $K^-\pi^+$ was in the $D^{\circ}$. Since it was known that $K^-\pi^+$ was about
3\%, this would have required a $\sim$40\% branching ratio for $F\to\eta\pi$.
Secondly, the production mechanism for $D$ decay had been shown to be mostly
associated production, where $\gamma p\to \overline{D}\Lambda_C X$, while in the
$F$ data the production was mostly $\gamma p \to F^+ F^- X$. Why should the $F$
production mechanism be so different? In any case the accepted $F$ mass now was
2020$\pm$10 MeV, having now been ``confirmed" by the CERN Omega data.

In July of 1980 one $F^+$ decay was detected in emulsion reactions from a
neutrino beam, in the here-to-fore unknown decay mode $\pi^+\pi^+\pi^-\pi^{\circ}$.
The mass was determined to be 2017$\pm$25 MeV, and lifetime of 1.4$\times
10^{-13}$s measured.\cite{ammar} Later in Sept. of that year another neutrino
emulsion experiment~\cite{ushida} measured a lifetime with two
events.\footnote{These
experiments prompted Lou Hand later to state: ``The $F$ was the first particle
whose lifetime was measured before it was discovered."}

\begin{figure}[htb]
\centerline{\psfig{figure=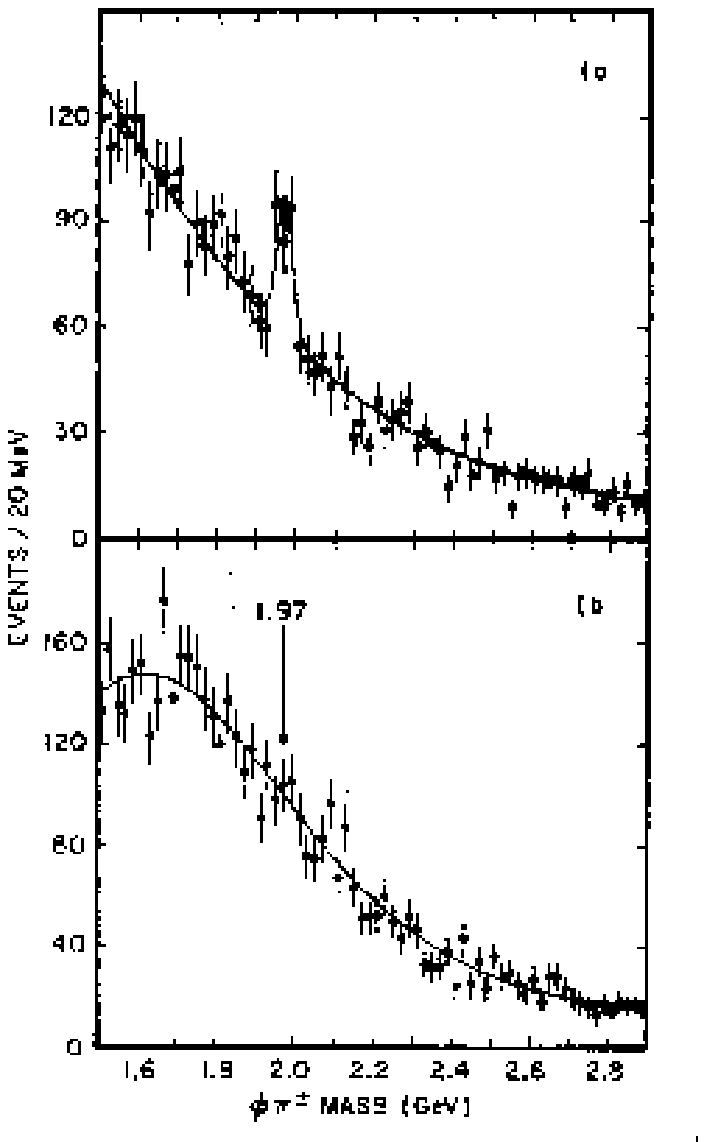,height=3.0in,angle=0.5}}
\vspace{-2mm}
\caption{(a) Mass spectra of  $\phi\pi^{\pm})$ from CLEO. (b) The $\phi$
candidates are chosen to be above or below the $\phi$ signal region. 
\label{fig:phipi_mass}}
\end{figure}

In 1983 the CLEO experiment presented results that showed strong evidence for
the $F$ at 1970$\pm$5$\pm$5 MeV. The evidence consisted of a mass peak 
containing 104$\pm$19 events in the
$\phi\pi^{\pm}$ decay mode shown in Fig.~\ref{fig:phipi_mass}, the helicity
distribution of the $\phi$, that showed the expected $\cos^2\theta$ decay
angular distribution and a $\sigma\cdot\cal{B}$ that
was 1/3 of the $\sigma\cdot\cal{B}$ for $D^{\circ}\to K^-\pi^+$.\cite{CLEO}
I can assure you that the CLEO collaboration was not easily persuaded by myself
and Yuichi Kubota that our results were right because they contradicted the
previous experiments shown above. We were forced to go through experiment by
experiment and detail what might have gone wrong. (Thus the material for this
section was created.) The CLEO result was quickly confirmed by
TASSO,\cite{TASSO} ACCMOR,\cite{ACCMOR} and
ARGUS experiments.\cite{ARGUS}
The Particle Data Group subsequently chose to rename the $F$ as the $D_s$, a logical
choice.

What lessons are we to learn from this story? DASP based their ``discovery" on
the obervation of increased $\eta$ production at 4.42 GeV and a handful of
events where one $F$ was reconstructed in $\eta\pi^{\pm}$ and the other not
reconstructed. Presumably they searched many energies and several final states
including $F^+F^-$ and $F^{*+}F^{*-}$, reporting a signal in only one case.
However, the statistical significance is often viewed by not considering all
the searches that yielded no signal.

The CERN Omega spectrometer results all were of very marginal accuracy and didn't
fit together very well. The $F$ production mechanism was completely different
than that for $D$'s, the rates for $\eta\pi^{\pm}$ were more than 7 times
larger than for $\phi\pi^{\pm}$, yet these discrepancies were never addressed.
We note that none of their results for the relative branching ratios are
consistent with current measurements.\cite{PDG} For
example, $\eta\pi^{\pm}$ is about half of $\phi\pi^{\pm}$ and $\phi\rho^{\pm}$ is
twice $\phi\pi^{\pm}$. Apparently their data showed marginal signals at 2020
MeV and the DASP result was sufficient to push them over the edge in giving
credibility to their fluctuations. 

The neutrino experiments now had the DASP and CERN Omega results to fall back
on. Doubtless if they hadn't believed the 2020 MeV value for the mass they may
not have shown their results; on the other hand, they could have been more
conservative. 

The DASP, Omega and neutrino results satisfy Langmuir's first criterion: ``The
maximum effect that is observed is....of barely detectable intensity...." Even
more so the second criterion is fully satisfied, especially by the Omega result: ``The effect is of a magnitude
that remains close to the limit of detectability, or many measurements are
required because of the low statistical significance of the results." 
The neutrino experiments satisfied his third criteria, ``Claims of great
accuracy," in that they measured lifetimes! Criteria (4) and (5) did not come
into play although (4) should have been invoked to explain production yields
and relative branching ratios. Criterion (6), ``Ratio of supporters to
critics ..." was funny in that most people believed the DASP and Omega results
until the CLEO result came out and then suddenly no one believed them.

\section{Conclusions}
Be suspicious of new results. Think them over and if they don't make sense then
doubt them. It doesn't mean they are wrong, just not proven. Sometimes it's
difficult to know when something is right or wrong.

It is also difficult to find out exactly what went wrong unless you are
directly involved with an experiment or had the opportunity to visit and
question as Langmuir had. Even Langmuir found it difficult to figure out
the process.
From Langmuir: ``I don't know what it is. That's the kind of thing that
happens in all these. All the people who had anything to do with these things
find that when you're through with them some things are inexplicable.
You {\it can't} account for Bergen Davis saying that they didn't calculate
those things from the Bohr theory, that they were found by empirical methods
without any idea of the theory. Barnes made the experiments, brought them in to
Davis, and Davis calculated them up and discovered all of a sudden that they
fit the Bohr theory. He said Barnes didn't have anything to do with that. Well,
take it or leave it. How {\it did} he do it? It's up to you to decide. I can't
account for it. All I know is that there was nothing salvaged at the end, and
therefore none of it was ever right, and Barnes never did see a peak. You can't
have a thing halfway right."

As a final note, Prof. Roodman at this school described how some current
experiments, KTeV, Babar and Belle have ensured that the final answer to their
most important measurements is actually hidden from the data analyzers until
they are satisfied that all systematic checks have been
performed.\cite{Roodman} In my view this is a useful technique and should be
employed more often. Another method that has been employed is to have
different groups within a collaboration obtain their results independently.

Hopefully reviewing these painful lesson will help others avoid the same
pitfalls.

\section{Acknowledgements}
I would like to thank Tom Ferbel for showing me Langmuir's paper long ago.
Thanks to K. T. Mahanthappa, H. Murayama and J. Rosner for organizing a
very interesting school and inviting me to participate. Ray Mountain and
Jon Rosner helped greatly by carefully reading and editing this paper.

\end{document}